\documentstyle[12pt]{article}
\newcommand{\be}{\begin{equation}}
\newcommand{\ee}{\end{equation}}

\newcommand{\nc}{\newcommand}
\nc{\barr}{\begin{eqnarray*}}
\nc{\ear}{\end{eqnarray*}}
\nc{\ol}{\overline}
\nc{\pa}{\partial}
\nc{\nb}{\nabla}
\nc{\ra}{\rightarrow}
\nc{\lw}{\longrightarrow}
\nc{\Lw}{\Longrightarrow}
\nc{\gm}{\gamma}
\nc{\ep}{\epsilon}
\nc{\Om}{\Omega}
\nc{\om}{\omega}
\nc{\al}{\alpha}
\nc{\bt}{\beta}
\nc{\Gm}{\Gamma}
\nc{\Lm}{\Lambda}
\nc{\lm}{\lambda}
\nc{\sm}{\sigma}
\nc{\ro}{\rho}
\nc{\dl}{\delta}
\nc{\Dl}{\Delta}
\nc{\vr}{\varphi}
\nc{\ov}{\over}
\nc{\th}{\theta}
\nc{\ha}{{1\ov 2}}
\nc{\lp}{\bigtriangleup}

\newcommand{\ba}{\begin{eqnarray}}
\newcommand{\ea}{\end{eqnarray}}
\newcommand{\eqn}[1]{\label{#1}\end{equation}}
\nc{\dg}{\dagger}
\nc{\lf}{\left}
\nc{\rt}{\right}

\newcommand{\Tr}{{\rm Tr}\,}
\newcommand{\tr}{{\rm tr}\,}
\newcommand{\beq}{\begin{equation}}
\newcommand{\eeq}[1]{\label{#1}\end{equation}}
\newcommand{\bea}{\begin{eqnarray}}
\newcommand{\eea}[1]{\label{#1}\end{eqnarray}}
\renewcommand{\Re}{{\rm Re}\,}
\renewcommand{\Im}{{\rm Im}\,}
\begin{document}
\baselineskip 18pt
\begin{titlepage}
\hfill  hep-th/9708119, NYU-TH-97/08/02, CPHT-S554-0897
\begin{center}
\hfill
\vskip .4in
{\large\bf Bound States at Threshold in Supersymmetric Quantum Mechanics}
\end{center}
\vskip .4in
\begin{center}
{\large Massimo Porrati}\footnotemark
\footnotetext{e-mail massimo.porrati@nyu.edu}
\vskip .1in
{\em CPHT Ecole Polytechnique\\F-91128 Palaiseau CEDEX, France\\ and\\
Department of Physics, NYU, 4 Washington Pl.,
New York, NY 10003, USA\footnotemark\footnotetext{Permanent address.}\\ and\\
Rockefeller University, New York, NY 10021-6399, USA}
\vskip .1in
and
\vskip .1in
{\large Alexander Rozenberg}\footnotemark
\vskip .1in
{\em Department of Physics, NYU, 4 Washington Pl.,
New York, NY 10003, USA}
\footnotetext{e-mail sasha.rozenberg@nyu.edu}
\end{center}
\vskip .4in
\begin{center} {\bf ABSTRACT} \end{center}
\begin{quotation}
\noindent
We propose a general method that allows to detect the existence of 
normalizable ground states in
supersymmetric quantum mechanical systems with non-Fredholm spectrum. 
We apply our method to show the
existence of bound states at threshold in two important cases: 1) the quantum
mechanical system describing H-monopoles; 2) the quantum mechanics of D0
branes.
\end{quotation}
\vfill
\end{titlepage}
\eject
\section{Introduction}
In recent times, our knowledge of non-perturbative string theory has greatly
progressed. A key element in this progress has been our ability to
identify certain quantum mechanical (or field-theoretical) systems which
describe non perturbative states in string theory. More precisely, BPS states
in a given string model are often put in one-to-one correspondence with
the ground states of some associated supersymmetric system.
Two examples are paramount:
the quantum mechanical reduction of a 6-d, N=1 supersymmetric (8 real
supercharges) field theory with
one Abelian vector multiplet, and a charged hypermultiplet, and the quantum
mechanical reduction of a d=10, N=1 supersymmetric (16 real supercharges)
$SU(N)$ gauge theory. The first system describes essentially an
H-monopole~\cite{W1} of the SO(32) superstring, while the second describes
$N$ D0 branes. In both cases
string duality arguments predict the existence of exactly one bound state at
threshold for these systems, since string dualities map the BPS states they
describe into Kaluza Klein modes of the graviton.
To check the existence of these ground states at threshold directly in the QM
model is thus a powerful check of the correctness of the duality hypothesis;
the existence of a normalizable bound state for any $N$ in the D0 brane
system is also essential for the consistency of the M(atrix) theory proposal
of ref.~\cite{BFSS}.

The problem in counting the ground states arises from the fact that they are at
threshold~\footnotemark
\footnotetext{For the D0 brane, this has been proven rigorously 
in~\cite{dWN1,dWN2}.}. This is because in the QM models described above, 
the bosonic
coordinates take value in a non-compact manifold, and the potential energy
has zero-energy valleys extending to infinity.

A possible way of counting the number of normalizable ground states (better,
the difference between the multiplicity of bosonic and fermionic ground
states), without explicitly
solving the Schr\"odinger equations is to use the Witten index
$W[\beta]=\Tr (-)^F \exp(-\beta H)$~\cite{W2}, and to notice that when the
low-energy spectrum is not too pathological, and
$\tr(-)^F\exp(-\beta H)$ is trace-class, the following identity holds:
$n_B-n_F=\lim_{\beta\rightarrow \infty}W[\beta]$. Here, $\Tr$ denotes the
trace over the whole Hilbert space while $\tr$ is the trace over the
(finite-dimensional) fermionic Hilbert space.
Moreover, when the Hamiltonian of a supersymmetric QM has a continuous
spectrum without a gap at zero energy, the usual arguments that ensure the
invariance of the Witten index~\cite{W2} have to be carefully re-examined. In
refs.~\cite{SS1,SS2}, 
such analysis has been performed for the
H-monopole, and for the D0-brane Hamiltonian with $N=2$ (see also~\cite{Y}).
The result of refs.~\cite{SS1,SS2,Y} is
that in both cases $\lim_{\beta\rightarrow\infty}W[\beta]=1$.

In this paper, using a different method, we find that $n_B-n_F=1$ in the
H-monopole system. We also find that the same result holds for D0 branes in
9 dimensions, provided that  three-dimensional D0 branes have no bound
states\footnotemark.
\footnotetext{In ref.~\cite{FH} it was proven rigorously that no bound states
exist for D0 branes in two dimensions. An argument due to S. Shenker shows that
a similar result should also hold in three dimensions, see also~\cite{dW}.}
Our method complements the arguments of refs.~\cite{SS1,SS2} since
it proves the existence of D0 brane
bound states with $N$ any prime number.
Moreover, it appears to be
straightforwardly generalizable to other, more complicated
QM systems relevant to non-perturbative string theory. One chief example is
the QM model of ref.~\cite{KT}, which is the reduction of a 4-d N=1
supersymmetric theory, and describes a configuration of four three-branes
corresponding to a BPS black hole with nonzero entropy.

To completely check the prediction of duality, one should find a vanishing 
theorem proving the uniqueness of the bound states. Unfortunately, such a
theorem does not exist yet in the literature. On the other hand, progress 
towards proving the {\em absence} of D0 brane bound states in three dimensions
were recently made in~\cite{H1}.

The main idea of this paper is the following. In a supersymmetric QM with at
least two (real) supercharges, there exists a {\em real} superpotential
function $W$~\cite{F}. A change in the superpotential, $w$, induces a
change in the supercharges as follows:
\beq
Q\rightarrow e^{-w}Qe^{w}\equiv Q_{w}, \;\;\;
\bar{Q}\rightarrow e^{w}\bar{Q}e^{-w}\equiv \bar{Q}_{w}.
\eeq{1}
If the bosonic fields take value in a compact space, or if $w /W$ goes
to zero at infinity in field space, the $L_2$ cohomologies of $Q$ and
$Q_{w}$ coincide, and therefore the number of normalizable ground
states of the QM with superpotential $W+w$ is the same as for
superpotential $W$~\cite{W2}. When the field space is non-compact, and the
potential obtained from $W$\footnotemark  has zero-energy valleys,
\footnotetext{Together with appropriate D-terms~\cite{F}.}
the argument does not work, generically: the local cohomologies of
$Q$ and $Q_{w}$ coincide, but the $L_2$ ones do not. Typically, one of the
two cohomologies is normalizable while the other is not.
In this paper, we show that, nevertheless, in the case of the H-monopole and D0
brane, it is possible to find an appropriate $w$ which acts as a
``dam'' for the valleys, lifting the flat directions in the potential, while
maintaining the normalizability properties of the cohomology. This special
perturbation allows us to reduce the computation of the index $n_B-n_F$ to
computing the Witten index of a system with mass gap and isolated minima in the
scalar potential. Such index can then be computed in the semiclassical
approximation.

The paper is organized as follows. In Section 2, we describe more precisely the
cohomology argument explained above. In Section 3, we find the
cohomology-preserving perturbation $w$ for the H-monopole QM, while
the perturbation for the D0 brane QM is discussed in Section 4, together with
an argument showing that no bound states exist in $d<5$.
Section 5 contains
our conclusions and a discussion of possible generalizations.
Some technical results are collected in the Appendices.
Appendix A contains a discussion about the validity of the adiabatic
(Born-Oppenheimer)
approximation used in this paper, and an argument to justify it. Appendix B
discusses some aspects of the semiclassical approximation in supersymmetric
QM, which are used in Section 4. The eigenvalues of the fermion mass matrix
of the perturbed D0 brane at the stationary point are computed in Appendix C,
together with some relevant properties of the Fock vacuum associated to
that mass matrix.

\section{The Cohomology Argument}
In this section we provide some criteria for the existence of the normalizable
supersymmetric ground states of a Hamiltonians with flat directions. These
criteria are based on the properties of the ground states of the perturbed
Hamiltonian $H_w\approx\ha\{Q_w,\bar{Q}_w\}$. As it was mentioned above,
despite the fact that the local cohomologies of the operators $Q$ and $Q_w$ are
equivalent, the mere
existence of normalizable ground states of the Hamiltonian $H_w$ does not prove
that they exist also in the original Hamiltonian $H_0$. However, they do exists
if additional properties are fulfilled. 
We summarize our method in the following
\newtheorem{guess}{Lemma}
\begin{guess}
There exists a normalizable ground state of the Hamiltonian $H_0$ if:
\begin{description}
\item[i)] The Witten index of the Hamiltonian $H_w$ is not zero.
\item[ii)] $\psi_w^{\pm}\equiv e^{\pm w}\phi_w$ are normalizable, 
where $\phi_w$ is a normalizable ground state of the perturbed Hamiltonian 
$H_w$. 
\end{description} 
\end{guess}
The condition (i) ensures the existence of a normalizable solution of the
perturbed Hamiltonian $H_w$. 

The functions $\psi^\pm_w$ are cohomology representatives of the operators 
$Q, \bar{Q}$, since they obey:
\beq
Q\psi_w^+=\bar{Q}\psi_w^-=0.
\eeq{madd1}
 On cohomological grounds we may write:
\bea
\psi_w^+&=&\al+\gamma,\;\;\; \psi_w^-=\bar{\al}+\bar{\gamma},\nonumber
\\
\gamma &\in& \overline{{\cal I}\,Q}, \;\;\;\alpha\notin\overline{{\cal I}\,Q},
\nonumber \\
\bar{\gamma}&\in& \overline{{\cal I}\,\bar{Q}},\;\;\; \bar{\alpha}\notin
\overline{{\cal I}\,\bar{Q}}.
\eea{madd8}
Here $\overline{{\cal I}\,Q}$ is the closure in $L_2$ of the image of $Q$ etc.
If both $\psi^+_w$ and $\psi^-_w$ are $L_2$, then so are $\alpha$ and 
$\bar{\alpha}$. 
By the definition of $\gamma$ and  $\bar{\gamma}$, there exist two sequences 
of $L_2$ states, $\beta_n$, $\bar{\beta}_n$, such that 
\beq
\gamma =\lim_{n\rightarrow \infty}Q\beta_n, \;\;\;
\bar{\gamma}=\lim_{n\rightarrow \infty}\bar{Q}\bar{\beta}_n .
\eeq{clos}
 
By computing the scalar product $(\psi^+_w,\psi^-_w)$ we find
\beq
1=(\psi_w^+,\psi_w^-)=\lim_{n\rightarrow \infty}(\alpha + Q\beta_n,\psi^-_w)=
(\alpha,\bar{\alpha}) + \lim_{n\rightarrow \infty}
(\beta_n, \bar{Q}\psi^-_w)=(\alpha,\bar{\alpha}).
\eeq{co1}
This equation, which is well defined when $\alpha$, $\beta_n$ etc. are in 
$L_2$, implies that $\alpha$ and $\bar\alpha$ cannot be zero.
Notice that this result  follows only if {\em both} $\psi^\pm_w$ are $L_2$.

The non-triviality of the cohomologies ensures the existence of a normalizable
ground state of $H_0$. This is a well known fact in mathematics; for 
completeness, we give here below a short proof of this result.

Define two orthogonal projectors, 
$P_n$, $R_n$, such that $P_n + R_n =1$. 
$P_n$ is the spectral projector associated to $H_0$, projecting
over eigenstates with energy $E\geq 1/n$, while $R_n$ projects 
over states
with energy less than $1/n$. For all $n > 0$, 
$P_n\alpha$ can be written as $Q\beta_n$. 
By defining $\beta_n=\bar{Q}H^{-1}P_n
\alpha$ (see also ref.~\cite{W2}); $\beta_n$ is $L_2$ since the operator
$\bar{Q}H^{-1}P_n$ is bounded. Analogously, 
$\bar{\beta}_n\equiv P_n\bar{\alpha}$ is
$\bar{Q}$ exact. Moreover, Since $\alpha$ is $L_2$, and ${\cal I}R_n \subset
{\cal I}R_m$ for $n>m$, the sequence $\alpha_n\equiv R_n\alpha$ is Cauchy;
therefore, it converges to a state 
$\tilde{\alpha}=\lim_{n\rightarrow \infty}\alpha_n$ which, by construction,
is normalizable and obeys $H_0\tilde{\alpha}=0$. The state
$\tilde{\alpha}$ cannot be
zero because of eq.~(\ref{co1}) and the definition of $\alpha$ 
given in eq.~(\ref{madd8}).

\noindent Q.E.D.

As we just noticed, if only one of the $\psi_w^\pm$ is normalizable, it may
happen that $(Q\beta_n, \psi^-_w)\neq 
(\beta_n,\bar{Q}\psi^-_w)=0$, so that the cohomology 
may still be empty. 
To illustrate this phenomenon, let us
consider a simple example: a free supersymmetric particle in one dimension,
which is known to have no normalizable ground states. Take a perturbing
superpotential $w=\ha x^2$. 
The perturbed problem has a zero energy normalizable
solution $\phi_w=\exp(-\ha x^2)$, so that $\psi_w^-$ is normalizable, while
$\psi_w^+$ is not. What actually happens in this example, is that $\psi_w^-$ is
cohomologically trivial, i.e. there exists $\chi$ s.t. $\psi_w^-=Q\chi$.

Note that the argument in the lemma cannot be reversed: an infinite norm of the
$e^{\pm w}\phi_w$ does not imply non-normalizability for $\phi_0$.

The Hamiltonian $H_w$ should be chosen in such a way that avoids the typical
problems associated with flat directions. Particularly simple for the
analysis is the case when the potential $V_w$ of the supersymmetric Hamiltonian
$H_w$ has only one isolated minimum. 
Then, since the Witten index is nonzero, this system
is guaranteed to have (at least) a normalizable ground states of zero energy, 
i.e. there exists (at least) a state $\phi_w$ s.t. $H_w\phi_w=0$, and 
$|\!|\phi_w|\!|<\infty$.

Given that the perturbing potential is a non-singular function,  and that the
function $\phi_w$ is square integrable, the normalizability of the
$\psi^{\pm}_w$ has to be checked only at infinity, since the integral over a
ball of radius $R_0$ yields:
\beq
|\!|\psi^{\pm}_w|\!|^2_{\downarrow R_0}\equiv \int_{|{\bf R}|<R_0}\,d{\bf R}
\lf|\psi^{\pm}_w\rt|^2\leq\exp\lf(2\,\mbox{sup}_{|\bf R|<R_0}\,w\rt)\,
|\!|\phi_w|\!|^2_{\downarrow R_0}<\infty.
\eeq{madd2}

\section{The H-Monopole QM}
The quantum mechanical system that describes the H-monopoles (better, the
``missing states'' in the problem~\cite{W1}) can be thought of as the
dimensional reduction of an N=2 field theory in four dimensions, whose
degrees of
freedom are an Abelian vector multiplet $V$, a neutral scalar multiplet $X$,
and two charged chiral multiplets, $\Phi^\pm$, of charge $\pm 1$ with respect
to the $U(1)$ gauge group.
The kinetic terms are canonical, while the superpotential is:
\beq
W=\Phi^+\Phi^- X.
\eeq{m1}
The scalar potential $V$ is given by the quantum-mechanical reduction of
the standard N=1 formula, namely:
\beq
V=W_i W^*_i + {1\over 4}D^2 + (|\phi^+|^2 + |\phi^-|^2)A^2_\mu.
\eeq{m2}
Here, $\phi^\pm$ etc. denote the scalars of the chiral multiplets;
$W_i$ denotes the derivative of the superpotential with respect to the $i$-th
scalar, and $A_{\mu}$, $\mu=1,2,3$ is the component of the vector in the
multiplet $V$ along the direction $\mu$.
The D-terms are given by
\beq
D= |\phi^+|^2-|\phi^-|^2.
\eeq{m3}

Following our general strategy, we add to this superpotential a perturbation
$w$, and we proceed to show that a) with the new superpotential $W+w$, there
exists a unique point where the (non-negative) scalar potential vanishes,
with non-vanishing Hessian, and thus its index
$n_B-n_F=1$; b) the perturbation lifts the moduli (valleys) $dW=0$ and does
not introduce new valleys; c) given a normalizable ground state of the
perturbed problem, $\phi_w$, the states $\psi^\pm=e^{\pm \Re w}\phi_w$ are
{\em normalizable} representatives of the cohomology of the original
supersymmetric quantum mechanics.

The perturbation we choose is
\beq
w= k X,
\eeq{m4}
with $k$ an arbitrary nonzero constant. This perturbation, being linear in
the coordinate $X$, does not change the fermionic part of the Hamiltonian. 

To prove point a), we have to solve the F-term equations $(W+w)_i=0$, and set
to zero the D-terms.
These equations read:
\beq
\phi^\pm x =0, \;\;\; \phi^+\phi^- + k =0,\;\;\; |\phi^+|^2-|\phi^-|^2=0.
\eeq{m5}
Here $x$ is the complex scalar in the supermultiplet $X$.
The D-term equation implies $|\phi^+|=|\phi^-|$, while the F-term equations
give (modulo a gauge transformation, obviously)
\beq
x=0,\;\;\; |\phi^\pm|=|k|^{1/2}; \;\;\; A_\mu=0,\; \mu=1,2,3.
\eeq{m6}
We do not need to compute the Hessian to see that it it nonzero. Indeed, since
$\phi^\pm\neq 0$, the 4-d N=1 model is in the Higgs phase: all fields are
massive. Upon dimensional reduction, this means that the Hessian of the
scalar potential $V$ is nonzero.

To prove point b) it is convenient to rewrite the scalar potential as
\beq
V={1\over 4}(|\phi^+|^2 + |\phi^-|^2)^2 +|k|^2 + 2\Re k\phi^+\phi^- +
 (|x|^2+A^2_\mu)(|\phi^+|^2 + |\phi^-|^2) .
\eeq{m7}
We want to prove that for $R^2\equiv
(|\phi^+|^2 + |\phi^-|^2 + |x|^2 + A^2_\mu)\rightarrow \infty$, the potential
attains its minimum value, $|k|^2$, along the valleys at $\phi^\pm=0$.
The minimization in $\phi^\pm$, keeping $ (|\phi^+|^2 + |\phi^-|^2)$ fixed,
gives $|\phi^+|=|\phi^-|$, and allows us to rewrite the potential as
\bea
V&=&{1\over 4}R^4\cos^4 \theta -|k|R^2\cos^2 \theta +R^4\cos^2 \theta
\sin^2 \theta + |k|^2, \nonumber \\ && R^2\cos^2 \theta \equiv
|\phi^+|^2 + |\phi^-|^2,\;\;\;  R^2\sin^2 \theta\equiv |x|^2+A^2_\mu.
\eea{m8}
For $R\rightarrow \infty$, the minimization in $\theta$ gives a
minimum at $\cos \theta =0$, i.e. along the flat directions of the unperturbed
model.

Point c) is crucial in showing that the perturbed problem has the same
cohomology as the unperturbed one.
To prove it, we need only study the asymptotic form of the ground state in the
perturbed problem, since $w$ is regular (and bounded) on any compact set.
To apply Lemma 1, one has to uncover the asymptotic form of the zero energy
solution in the perturbed problem at large distances. While unable to solve the
problem exactly, we may use the Born-Oppenheimer approximation in the
asymptotic
region. Indeed, as it follows from the form of the scalar potential, the
frequency of the oscillations in the directions transverse to the moduli
subspace $\phi^\pm=0$ is $\om_{trans.}\geq \ha R^2$, while the characteristic
frequency of the oscillations on the moduli is $\om_{moduli}\leq |k|$.
For $R$ large
enough, the separation into ``slow'' and ``fast'' modes holds.
 We can label the 5-d moduli space with the vector
\beq
{\bf x}\equiv(\Re x, \Im x, A_1, A_2, A_3),
\eeq{m9}
and define:
\beq
{\bf k}\equiv(\Re k, \Im k, 0,0,0).
\eeq{m10}
Along the flat directions, the wave function, $\Psi({\bf x})$, satisfies:
\be\label{equ}
\lf(-\ha\nabla_{\bf x}\nabla_{\bf x} +\ha \lf|{\bf k}\rt|^2\rt)
\Psi({\bf x})=0.
\ee
Re-writing this equation in spherical coordinates, and denoting by $\om$ the
angular variables, and by $r$ the radius $|{\bf x}|$, one obtains:
\be
\label{equ'}
\lf({d-1\ov r}{\pa\ov\pa r}+{\pa^2\ov\pa r^2}+{1\ov r^2}{\pa^2\ov\pa
\om^2}-k^2\rt)\Psi({\bf x})=0.
\ee
After separation of variables; eq.~(\ref{equ'}) gives, for the
radial part of the wave function, $\Psi({\bf x})=R(r)\Om(\om)$:
\be
\label{bes1}
\lf({d-1\ov r}{\pa\ov\pa r}+{\pa^2\ov\pa r^2}-\lf({{\bf L}^2\ov
r^2}+k^2\rt)\rt)R(r)=0,
\ee
where ${\bf L}^2$ is the square of the total angular momentum.
The asymptotic behavior of the solutions to eq. (\ref{bes1}) at large $r$ is:
\be
\label{sol2}
R(r)\stackrel{r\rightarrow\infty}{\longrightarrow}{\rm const}\,
{e^{-kr}\ov
r^{{d-1\ov 2}}}\lf(1+{\cal O}\lf({1\ov r^2}\rt)\rt).
\ee
The wave function in the transverse directions decays at least as fast as that
of a harmonic oscillator, and the perturbation $w$, being linear in
coordinates, cannot spoil the normalizability.

The normalizability of the ``slow'' modes functions $\psi^{\pm}({\bf
x})=e^{\pm\Re w}\Psi({\bf x})$ can be checked as follows:
\ba\label{int}
& &\int_{|{\bf x}|>r_0}d{\bf x}\,\lf|\psi^\pm({\bf
x})\rt|^2={\rm const}\int_{r>r_0}d{\bf x}{e^{-2|k|r\pm 2{\bf k}{\bf x}}\ov
r^{d-1}}\nonumber\\
&=&{\rm const}\,
\Om_{d-2}\int_{r=r_0}^{\infty}dr\, e^{-2|k|r} \int_{0}^{\pi}d\th
\sin^{d-2}\th \,e^{\pm 2|k|r\cos\th}=
{\rm const}\,\int_{r=r_0}^{\infty}{dr\ov r^{{d-1\ov
2}}}.
\ea
The integral in (\ref{int}) converges if $d>3$.
Since in the case of the H-monopole the dimension of the
moduli space is $d=5$, the criteria for the existence of the normalizable
supersymmetric ground state are satisfied.

Our argument proves that there always exists at least a supersymmetric ground
state in the H-monopole problem. Barring accidental degenerations, this ground
state should also be unique, in agreement with the predictions of S-duality.

\section{The D0 Brane QM}
The conjectured duality between the type IIA string theory and M
theory~\cite{W3} as well as the M(atrix) model conjecture~\cite{BFSS} require
that a system of $N$ D0 branes in Type IIA string theory has a unique bound
state at threshold for any $N$. In this section, we show that such a state
exists for any prime $N$. For $N=2$, the existence of
a zero-energy bound state has been proven in ref.~\cite{SS2}.

It is convenient to rewrite the D0 supersymmetric QM in a formalism that makes
manifest just four of its 16 supersymmetries, as we did for the H-monopole.
This is achieved by first reducing
the 10-d $SU(N)$ super Yang-Mills theory~\cite{BSS} 
--which describes $N$ D0 branes-- to
four dimensions, and then perform another dimensional reduction to the 1-d 
QM~\cite{CH}.
In the 4-d, N=1 language, the theory is made of a
vector superfield, $V$, and 3 chiral superfields, that we shall denote with
$\Phi_i$, $i=1,2,3$.

The bosonic degrees of freedom of the vector multiplet
are a 4-d vector
$A_\mu$ in the adjoint of the gauge group, and some auxiliary fields.

The bosonic degrees of freedom of the scalar multiplets are some auxiliary
fields and the complex scalars
$\phi_i$, also in the adjoint of the gauge group.

The model has canonical kinetic terms, and it is completely determined by
its superpotential, which is a holomorphic, gauge-invariant function:
\beq
W={1\over 6}
\epsilon_{ijk}\tr \Phi_i[\Phi_j,\Phi_k].
\eeq{m11}
Here $\tr$ denotes a trace over the gauge indices.
This choice of superpotential ensures that the model is invariant under 16
supersymmetries, and not just the four which are manifest in this formalism.

The scalar potential of the quantum mechanical model reads, in the $A_0=0$
gauge:
\beq
V = \left|{\partial W\over \partial \phi_i}\right|^2 +
\tr [\bar{\phi}_{\bar{\imath}},\phi_i][\bar{\phi}_{\bar{\jmath}},\phi_j]
 +\tr [A_\mu,\phi_i][A_\mu,\bar{\phi}_{\bar{\imath}}] +
{1\over 4}\tr [A_\mu,A_\nu]^2 .
\eeq{m12}
Here, $\mu, \nu= 1,2,3$.

\subsection{The Deformation}
To exploit the technique described in Section 2, we must first deform
the model by adding to it an appropriate superpotential $w$.
Our choice for $w$ is
\beq
w=-{1\over 2}m\tr\Phi_i^2,
\eeq{m13}
with $m$ a nonzero constant.
This perturbation has been introduced in ref.~\cite{VW}; in 4-d it
gives a mass $|m|$ to the chiral multiplets.

The zeroes of the perturbed potential are given by the solutions of the
following equations, modded out by the action of
the gauge group:
\bea
[A_\mu,A_\nu]&=&0,\;\;\; [A_\mu,\phi_i]=0,\;\;\; \mu=1,2,3, \label{m14} \\
D&=&[\bar{\phi}_{\bar{\imath}},\phi_i]=0, \label{m15} \\
W_k + w_k=0 &\Rightarrow &
[\phi_i,\phi_j]= m\epsilon_{ijk}\phi_k.
\eea{m16}
Eqs.~(\ref{m15},\ref{m16}) set to zero the D- and F-terms. This is equivalent
to solve eq.~(\ref{m16}) and mod out by the complexified gauge group. The
total Witten index is obtained by summing over the contributions to the index
of each such solution.
These equations have been studied in ref.~\cite{VW}, with the result that their
solutions are in one-to-one correspondence with the complex conjugacy classes
of $SU(2)$ into the gauge group $G$, that is, with the inequivalent
representations of $SU(2)$ into the fundamental of $G$.
There are three types of
representations to be considered, generically: the trivial ($\phi_i=0$), the
irreducibile, and some reducible ones. The trivial representation
is given by $\phi_i=0$. At this point the chiral multiplets have a 4-d ``mass''
$|m|$ that, upon dimensional reduction, implies that these modes have a nonzero
frequency, and thus do not contribute to the index.
This means that at $\phi_i=0$ the D0 brane system is effectively three
dimensional. If, as suggested in ref.~\cite{dW}, the 3-d D0 brane has no bound
states, then the $\phi_i=0$ minimum gives a null contribution to the index. In
the rest of the paper we will assume that this result holds;
later in this section we will report an argument supporting this assumption.

When the gauge group is $SU(N)$, and $N$ is prime, the reducible
representations break the gauge group to $U(1)^{N-1}$. The light degrees of
freedom parametrize an Abelian theory, which again gives a null
contribution to the index.

Finally, the only nonzero contribution to the index
comes from the unique irreducible representation of dimension $N$. It breaks
$SU(N)$ completely, implying that all degrees of freedom are massive, i.e. that
the minimum is at $A_\mu=0$ and that the Hessian determinant of the potential
is nonzero. To sum up, the irreducible representation gives an isolated minimum
contributing 1 to the index, while all other minima give no contribution.

\subsection{The $L_2$ Cohomology}
The argument above implies that $n_B-n_F=1$\footnotemark,
\footnotetext{Obviously, the definition of bosons and fermions in
supersymmetric QM is conventional, and it can be changed with a unitary
redefinition of the fermionic Fock vacuum.}
i.e. that in the perturbed problem with superpotential $W+w$ there exists at
least one normalizable ground state.

Next, following our general strategy outlined in Section 2,
we must show that it is possible to
associate to the perturbed ground state, $\phi_w$, two {\em normalizable}
representatives, $\psi^\pm_w$, of the unperturbed cohomologies $Q$ and
$\bar{Q}$.

As in the H-monopole case, we set
\beq
\psi^\pm= (\exp\pm \Re w )\phi_w.
\eeq{m19}

To study the normalizability of $\psi^\pm_w$ we can limit ourselves to the
asymptotic, large $\phi_i$ region, since for finite $\phi_i$ both $\phi_w$
and $\psi_w^\pm$ are smooth and obviously normalizable.

We want, first of all, to find the new ``valleys'' of the perturbed potential,
$V_w$,
for large $\phi_i$, i.e. to find its minima for fixed, large radius
$R=(\tr \phi_i \bar{\phi}_{\bar{\imath}})^{1/2}$.
This can be done by adding to the
potential a Lagrange multiplier, and looking for the stationary points
of
\bea
V_w+ \lambda(\tr \phi_i \bar{\phi}_{\bar{\imath}}-R^2) &=&
\left|{\partial (W+w)\over \partial \phi_i}\right|^2 +
\tr [\bar{\phi}_{\bar{\imath}},\phi_i][\bar{\phi}_{\bar{\jmath}},\phi_j]
 +\tr [A_\mu,\phi_i][A_\mu,\bar{\phi}_{\bar{\imath}}] +
{1\over 4}\tr [A_\mu,A_\nu]^2 +\nonumber \\
&& +\lambda(\tr \phi_i \bar{\phi}_{\bar{\imath}}-R^2),
\eea{m20}
in the limit $R\rightarrow \infty$.
Minimization of eq.~(\ref{m20}) for large $R$ shows that the valleys of the
perturbed potential are the same of the unperturbed one:
\beq
[\phi_i,\phi_j]=[\phi_i,\bar{\phi}_{\bar{\imath}}]=0,
\eeq{m21}
and that along the valleys $V_w=|m|^2 R^2$.

Our aim is to prove that the following integral converges:
\beq
\int_{\tr \phi_i \bar{\phi}_{\bar{\imath}}\geq R^2}d\mu
\psi^{\pm\, *}_w \psi^\pm_w .
\eeq{m23a}
Here $d\mu$ denotes the (flat) integration measure of all bosonic variables.
For $R \gg 1$, we can use again the adiabatic approximation, and reduce the
integral to
the estimate of the behavior of the adiabatic wave function on the
$9(N-1)$\footnotemark  moduli space given by eq.~(\ref{m21}).
\footnotetext{Here
we write all formulae for the gauge group $SU(N)$, of course.}

The asymptotic behavior of the wave function $\phi_w$ may be estimated as
follows.

On the moduli space, the Born-Oppenheimer wave function must satisfy
the equation
\beq
H\phi_w=0, \;\;\;
H=H_1+H_2,\;\;\; H_1= -\triangle_x,\;\;\; H_2= -
\nabla_{y}\nabla_{\bar{y}} +
|m|^2|y|^2  +|m|[\psi_\alpha^\dagger \psi_\alpha-3(N-1)].
\eeq{m24}
The effective Hamiltonian $H$ is just the reduction of the perturbed
D0 brane Hamiltonian to the moduli space. Here,
$\triangle_x$ is the $3(N-1)$-dimensional Laplacian acting on the bosonic
moduli
of the vector multiplet: $x\equiv (A_1, A_2, A_3)$. We parametrized the
moduli space of the chiral multiplets $\Phi_i$ by $3(N-1)$ superfields
$Y=y +\theta^\alpha \psi_\alpha$, $\alpha=1,2$.

The asymptotic behavior of $\phi_w$ is {\em generically} the same as that of
the Green
function $G=H^{-1}$, even though it may be better, in some exceptional cases.
By diagonalizing $H$ and denoting by $K=0,..,6(N-1)$ the
eigenvalue of the fermion number operator $\psi_\alpha^\dagger \psi_\alpha$
we find:
\beq
G(x,y,K|x',y', K')=\delta_{KK'}\sum_{n=0}^\infty
\int {d^{3(N-1)} p\over (2\pi)^{3(N-1)}}
{1\over p^2 + |m|(n+K)} e^{ip(x-x')}
\Phi_n(y)^*\Phi_n(y'),
\eeq{m25}
where the $\Phi_n(y)$ are the usual eigenstates of the $6(N-1)$-dimensional
harmonic oscillator with frequency $|m|$. By using the Schwinger
representation for the free propagator and integrating in $p$ we find:
\beq
G(x,y,K|x',y', K')={1\over 2}(2\pi)^{-3(N-1)/2}\int_{t=0}^\infty dt
t^{-3(N-1)/2} \exp(-|x-x'|^2/2t)
  \langle y,K | \exp(-t H_2) |y',K'\rangle.
\eeq{m26}
Eq.~(\ref{m26}) can be computed with a standard Euclidean functional integral,
which gives
\bea
G(x,y,K|x',y',K')&=&\delta_{KK'}
{1\over 2}(2\pi)^{-3(N-1)/2}\int_0^\infty dt
t^{-3(N-1)/2} \exp\{-|x-x'|^2/2t+|m|t
[3(N+\nonumber \\ && -1)- K]\}
\left(|m|\over \sinh |m|t\right)^{3(N-1)}
\exp\Big\{-{|m|\over \sinh |m|t}
[(|y|^2 +\nonumber \\ && + |y'|^2)\cosh |m|t -2\Re (y^*y')]\Big\}.
\eea{m27}

The ground-state wave function $\phi_w$, and thus $\psi^\pm_w$, is not an
eigenstate of the fermion number operator $\psi^\dagger_\alpha\psi_\alpha$, as
explained in more detail in Appendix B. Rather, $\psi^\pm_w$ is a
superposition of wave functions with different fermion numbers. Asymptotically
in $|y|$ one has:
\beq
\psi^\pm_w\approx \sum_{K=0}^{6(N-1)}c_KG(x,y,K|x',y',K)\exp(\pm \Re m y^2),
\eeq{m27aaa}
for some constant coefficients $c_K$.
The asymptotic decay rate of the Green function~(\ref{m27}) increases with
$K$, which implies that, for large $|y|$:
\beq
\psi^\pm_w\approx c_{K^o}G(x,y,K^o|x',y',K^o)\exp(\pm \Re m y^2),
\eeq{m27aa}
where $K^o$ is the smallest $K$ appearing in equation~(\ref{m27aaa}) with
a nonzero coefficient $c_{K}$.
To prove that the $\psi_w^\pm$ are normalizable we need only study the large
$\phi_i$ region, i.e. $|y| \gg |y'|$. We first integrate in $x$
and obtain
\bea
\int d^{3(N-1)}x |\psi_w^\pm(x,y)|^2 &\approx &
{|c_{K^o}|^2\over 2}
\int_0^\infty ds\int_0^\infty dt (t+s)^{-3(N-1)/2} \exp\{|m|(t+s)
[3(N-1)+ \nonumber \\ && -K^o]\}
\left(|m|^2\over \sinh |m|s\sinh |m|t \right)^{3(N-1)}
\exp[-|m|(\coth |m|s+ \nonumber \\ && +\coth |m| t)|y|^2 \pm 2\Re m y^2].
\eea{m27a}
We can restrict the region of integration to large values of $y$ by
restricting $y$ to lie outside of the hypercube ${\cal C}$, defined by
$ \Re y_i, \Im y_i\leq R \;\forall i$, with $R \gg |m|^{1/2}$. We find:
\bea
\int_{R^{6(N-1)}-{\cal C}} d^{6(N-1)}y  |\psi_w^\pm|^2 &\approx &
A\int_0^\infty ds\int_0^\infty dt
(t+s)^{-3(N-1)/2} \exp\{|m|(t+s) [3(N -1)- K^o]\}\times \nonumber \\ &&
\{\sinh^2 |m|s\sinh^2 |m|t[(\coth |m|s + \coth |m|t)^2 -4]\}^{-3(N-1)/2}
\times \nonumber \\ &&
\exp[-R^2|m|(\coth |m|s +\coth |m|t -2)],
\eea{m27b}
where $A$ is a finite, positive constant. The dangerous
region of integration for the norm of $\psi_w^\pm$ is where $s$ and $t$ are
both large. There, eq.~(\ref{m27b}) simplifies considerably, and one finds
\bea
\int_{R^{6(N-1)}-{\cal C}} d^{6(N-1)}y  |\psi_w^\pm|^2 &\approx &
A\int_M^\infty ds\int_M^\infty dt
(t+s)^{-3(N-1)/2}\exp\{|m|[3(N-1)/2-K^o](s+t)\} \nonumber \\ &&
\times[\cosh |m|(t-s)]^{-3(N-1)/2}+B =
\int_{2M}^\infty du \int_{-\infty}^{\infty}dv
u^{-3(N-1)/2}\times \nonumber \\ &&
\exp \{|m|[3(N-1)/2-K^o]u\}(\cosh|m|v)^{-3(N-1)/2} +B.
\eea{m27c}
Here $M\gg 1/|m|$, $u=s+t$, $v=s-t$,
and $B$ is another finite, positive constant.
As explained in Appendix C, the ground-state wave function is the completely
filled state in the Fock space generated by a subset of the oscillators
$\psi_\alpha$, $\psi^\dagger_\alpha$. Using this result, it is easy to compute
the fermion number $K^o$. Indeed, for each $2\leq l\leq 2j$ there exist
$2\times 3$ such oscillators, while for $l=1$ there are only two such
oscillators\footnotemark; therefore,
\beq
K^o=\sum_{l=2}^{2j}6 +2= 6(2j-1) +2 = 6N -10.
\eeq{m27d}
The integral in eq.~(\ref{m27b}) is finite if and only if $K^o>3(N-1)/2$.
This happens for $N>17/9$, and
therefore the cohomology representatives $\psi^\pm_w$ are $L_2$ for all
$SU(N)$ with $N$ prime.
\footnotetext{See Appendix C for notations. The
factor 2 comes from the index $\alpha=1,2$; the factor 3 comes from the index
$s_3=0,\pm 1$. For $l=1$ only $s_3=0$ is allowed.}

 By applying the method of Section 2, we
conclude that a system of $N$ D0 brane in 9 dimensions has at least a
zero-energy bound state for any prime $N$.

This conclusion, in agreement with expectations from string duality
and Matrix theory, holds if no bound states exist in the 3-d D0 brane system.
An argument which shows that this is the case has been formulated by S.
Shenker. It goes as follows. The asymptotic moduli
space can be partitioned into $N-1$ subregions in which the
VEVS of the scalars break $SU(N)$ to $SU(N-n)\times U(1)^n$, $n=1,..,N-1$.
Geometrically, this
corresponds to have $N-n$ D0 branes close to each other and $n$ far away.
Along this flat direction, the ground-state wave function approximately
obeys a free $dn$-dimensional Laplace equation, whose long-distance behavior
is generically given by the
Green function
$G(x,0)=|x|^{2-dn}$, $x\in R^{dn}$.
This function is square summable at large $x$ only if $nd>4$; for $n=1$ one
gets $d>4$.
This argument is incomplete (in some cases the
asymptotic behavior of the wave function can be better than that of
the Green function), but still rather compelling, especially when coupled
with the rigorous results of~\cite{FH}.

\section{Conclusions}
In this paper, we proposed a novel way of studying the existence of bound
states at threshold in supersymmetric QM systems. We applied our method to
the case of H-monopoles, and re-derived a known result, namely that there is
(at least) one normalizable bound state at threshold in the $U(1)$ theory
describing the ``missing'' H-monopoles~\cite{W1}.

We also studied a QM with
gauge group $SU(N)$ and 16 supercharges, that describes a system of $N$ D0
branes in nine dimensions.
Applying our method we found that this system possesses (at least) a
bound state at threshold for any prime $N$,
assuming that no bound states of D0 branes exist in three dimensions.
This result complements the one of
ref.~\cite{SS2}, in which the existence of a bound state was proven for $N=2$.
We believe that the restriction on $N$ is purely technical and not due
to any fundamental limitation of our method.

Our method uses the stability of the supercharge cohomology under appropriate
deformations of the superpotential. Compared with the technique of
refs.~\cite{SS1,SS2}, it has the advantage that it does not require the
computation of a massive multi-dimensional ``bulk'' integral, and the
delicate estimate of a boundary term.

Besides these technical points, we think that our method may be of interest
since it seems possible to extended it to the computation of the
index of more complicated (and less supersymmetric) QM models, such as the
QM describing 4-d BPS black holes described in~\cite{KT}.

Finally, let us emphasize that our method always determines wave functions
which are in the same cohomology class of the true ground state, rather than
the ground state itself. This means, in particular, that our method does not
determine
the true asymptotic behavior of the ground-state wave function, which
may be very different from that of the cohomology representatives.
\vskip .2in
\noindent
{\bf Acknowledgments}\vskip .1in
\noindent
We would like to thank J. Cheeger for useful discussions, and M. Stern for
useful comments on the manuscript.
M.P. is supported in part by NSF grants no. PHY-9318781, PHY-9722083; A.R.
is supported in part by a Margaret and Herman Sokol Research Fellowship.
A.R. would like to thank the LPTHE at the University of Paris VI for its kind
hospitality and support.
\section*{Appendix A}
\renewcommand{\theequation}{A.\arabic{equation}}
\setcounter{equation}{0}
The deformed Hamiltonian $H_w$ may still turn out to be too complicated to be
solved exactly. However, we are interested only in the asymptotic behavior of
the solutions, which can be found in the Born-Oppeheimer approximation.

Typically, since the perturbation is small compared to the original Hamiltonian
away from the subspace of flat directions, and the valleys in the original
problem become increasingly narrow, the frequency of the oscillations
transverse
to the original valleys is very large compared to the frequency along the
valleys. Given that, one separates the variables into ``slow'' $\xi$, and
``fast'' $\eta$, and introduces the corresponding Hamiltonians:
\be\label{hh}
H(\xi,\eta)=H_1(\xi,\eta)+H_2(\xi)
\ee
The $H_1$ piece depends on the slow variables $\xi$ only parametrically,
i.e. it contains no derivatives with respect to $\xi$.

The eigenfunctions of the full Hamiltonian satisfy
\be\label{anz2}
(H_1+H_2)\Phi(\xi,\eta)=E(\xi,\eta)\Phi(\xi,\eta),
\ee
and in adiabatic approximation are factorized as:
\be
\label{anz}
\Phi(\xi,\eta)=\sum \Xi(\xi)\Psi(\xi,\eta).
\ee
The $\Psi(\xi,\eta)$ are the eigenfunctions of $H_1$:
\be\label{add4}
H_1(\xi,\eta)\Psi(\xi,\eta)=E'(\xi)\Psi(\xi,\eta).
\ee
Both Hamiltonians $H_{1,2}$ are supersymmetric and hence non-negative, and
for that reason in order to find the ground state of (\ref{anz2}), we have to 
look for zero energy solutions of the corresponding
Hamiltonians. The existence of a solution to (\ref{anz2}) is implied by
the non-vanishing of the Witten index, and we are only concerned about the
asymptotic form of the solutions.
We want to show that the ground state $\Psi^{\lf(0\rt)}(\xi,\eta)$ of 
eq.~(\ref{add4}), and the ground state $\Xi^{\lf(0\rt)}(\xi)$, solution of 
\be\label{add5}
H_2\Xi^{\lf(0\rt)}(\xi)=0,
\ee
give an adequate asymptotic approximation of the ground state wave 
function $\Phi^{\lf(0\rt)}(\xi,\eta)$.

Recall that the asymptotic form of a solution of our
Fredholm differential operators is not affected by terms of order
$(\mbox{distance})^{-2}$. The zeroth order terms give the leading exponent,
while the terms of order $(\mbox{distance})^{-1}$ provide the subleading
power correction. Therefore, when solving eq.~(\ref{anz2}), one may
neglect corrections of order $(\mbox{distance})^{-2}$ in the equation for
$\Phi(\xi,\eta)$.

Since $H_2(\xi)=\ha\{Q_\xi,\bar{Q}_\xi\}$ is a supersymmetric Hamiltonian, one
has $Q_\xi\Xi^{\lf(0\rt)}(\xi)=\bar{Q}_\xi\Xi^{\lf(0\rt)}(\xi)=0$. Therefore:
\be
\label{boe1}
\lf(H_1(\xi,\eta)+H_2(\xi)\rt)\Xi^{\lf(0\rt)}(\xi)\Psi^{\lf(0\rt)}(\xi,\eta)
=-\Xi^{\lf(0\rt)}(\xi)\lf(\bigtriangleup_{\xi}\Psi^{\lf(0\rt)}(\xi,\eta)\rt).
\ee
Multiplying the r.h.s of eq.~(\ref{boe1}) by 
$\Psi^{\lf(0\rt)^*}(\xi,\eta)$, and integrating over $\eta$, one obtains:
\be\label{boe2}
\Xi^{\lf(0\rt)}(\xi) N^2(\xi)\int \, d\eta
\,\Psi^{\lf(0\rt)^*}(\xi,\eta)\lf({\pa^2\ov\pa\xi^2}+
{d_{\xi}-1\ov\xi}{\pa\ov\pa\xi}\rt) \Psi^{\lf(0\rt)}(\xi,\eta).
\ee
The variable $\xi$ is treated as a parameter in the equation
$H_1(\xi,\eta)\Psi^{\lf(0\rt)}(\xi,\eta)=0$. The function 
$\Psi^{\lf(0\rt)}(\xi,\eta)$ is of the form
\beq
\Psi^{\lf(0\rt)}(\xi,\eta)=N(\xi)\,{\cal F}(\xi^\al\eta),
\eeq{madd4}
where $N$ is a normalization factor, and $\al(\eta)\geq\al_0>0$ depends on the
particular potential. For a harmonic oscillator, $\al$ is of course
$\ha$. After a change of variables to $\mu\equiv\xi^\al\eta$, 
eq.~(\ref{boe2}) takes the form:
\be\label{boe3}
\Xi^{\lf(0\rt)}(\xi) N'^2\int \, d\mu \,{\cal
F}^*(\mu)\lf({\pa^2\ov\pa\xi^2}+{d_{\xi}-1\ov\xi}{\pa\ov\pa\xi}\rt){\cal
F}(\mu),
\ee
the where $N'$ is just a number:
\beq
{1\ov N'^2}=\int \, d\mu \lf|{\cal F}(\mu)\rt|^2.
\eeq{madd5}
Due to the definition of $\mu$, each differentiation in eq.~(\ref{boe3}) 
brings down one
power of $\xi$, so that eq.~(\ref{boe2}) becomes:
\be\label{add6}
{\Xi^{\lf(0\rt)}(\xi)\ov\xi^2}\times\lf(\mbox{integral independent
of}\,\xi\rt).
\ee
Consequently, $\Phi^{\lf(0\rt)}(\xi,\eta)=\Xi^{\lf(0\rt)}(\xi)
\Psi^{\lf(0\rt)}(\xi,\eta)$, obtained from eqs.~(\ref{add4},\ref{add5}), 
gives the ground state of eq.~(\ref{hh}) up to order $\xi^{-2}$.

\section*{Appendix B}
\renewcommand{\theequation}{B.\arabic{equation}}
\setcounter{equation}{0}
In this appendix, we want to study the semiclassical approximation for a
supersymmetric quantum mechanics.

In this approximation, the zero-energy wave function takes the form
\beq
\Psi(x)=\exp[-S(x)/\hbar] F(x) + O(\hbar).
\eeq{mb1}
Here $x$ denotes all bosonic variables, while $F(x)$ is a map from $x$ into the
fermionic Fock space.
The zero-energy ground state of a supersymmetric system obeys a Schr\"odinger
equation which reads, schematically:
\beq
\left[-{\hbar^2\over 2}\nabla^2_x + V(x) + \hbar M(x) \right] \Psi(x)=0.
\eeq{mb2}
$V(x)\geq 0$ is the scalar potential, which by assumption has an isolated,
non-degenerate zero at $x=0$, while the ``fermion mass'' term $M(x)$ is
a map from $x$ into the linear operators of the Fock space, i.e. it is an
$x$-dependent finite-dimensional matrix.
By substituting the ansatz eq.~(\ref{mb1}) into the Schr\"odinger equation and
expanding in powers of $\hbar$, we
find the equations:
\beq
{1\over 2}\nabla_x S\nabla_x S - V=0, \;\;\;
\left[(\nabla_x S)\nabla_x + M(x) + {1\over 2}(\nabla^2_xS)\right]F(x)=0.
\eeq{mb3}
The first equation is the familiar Hamilton-Jacoby equation of a bosonic system
with potential $-V(x)$, while the second determines the vector $F$ in the Fock
vacuum. By denoting with $N(x)$ the matrix $M(x) + I(\nabla_x^2S)/2$
($I$ is the
identity in the Fock space), we can easily write down the solution for $F(x)$
as
\beq
F(x)=\lim_{\tau\rightarrow -\infty}T\exp[-\int_{\tau}^{t(x)} N(x(s))] F(0).
\eeq{mb4}
Here, $T$ denotes the ordering of the matrices $N$ in the parameter $s$.
$x(s)$ is the zero-energy classical trajectory approaching $x=0$ at
$s\rightarrow -\infty$, and reaching $x$ at $s=t(x)$. A general property of a
supersymmetric system is that the eigenvalues of $N(x)$ are non-negative near
$x=0$. In
particular, $N(0)$ has a unique zero eigenvalue, $|0\rangle$,
which may be used to define
the Fock vacuum. Notice also that the trajectory $x(s)$ in eq.~(\ref{mb4})
spends an infinite amount of time near $x=0$. This fact, together with the
semipositive-definitness of $N$, implies that all components of $F(0)$ are
projected out by the operator $\lim_{\tau\rightarrow -\infty}
T\exp[-\int_{\tau}^{t(x)}dsN(x(s))]$, except the one parallel to the
Fock vacuum $|0\rangle$.
Therefore, we can choose as initial condition in eq.~(\ref{mb4})
$F(0)={\rm const}\, |0\rangle$. Since $N(x(s))$ has fermion number zero,
$F(x)$ has the same fermion number as $F(0)$, i.e. it is
proportional to the Fock vacuum {\em defined by} $N(0)$.
This last remark is important, since
by using another $N(x)$ one may define another Fock vacuum, which may not
coincide with the previous one. In terms of the new Fock vacuum,
the old one (generically) is not even an eigenstate of the new fermion number
operator. For the case of the D0 brane, this was explicitly shown in 
ref.~\cite{dWN1}.

\section*{Appendix C}
\renewcommand{\theequation}{C.\arabic{equation}}
\setcounter{equation}{0}
The second derivative of the superpotential of the perturbed D0 brane system
is:
\beq
(W+w)''_{ij}(\phi)=\epsilon_{ilj}\phi_l-m\delta_{ij}I,
\eeq{c1}
where $I$ is the identity matrix acting (as $\phi_l$ does) on the adjoint of
$SU(N)$. It determines the frequencies of the fermionic oscillators in the
chiral multiplets, since the fermionic Hamiltonian is
\beq
H_F=\epsilon_{\alpha\beta}\psi_\alpha^i (W+w)''_{ij}(\phi)\psi_\beta^j+ {\rm
h.c.}
+{\rm gaugino \; terms}, \;\;\; \alpha,\beta=1,2.
\eeq{c1a}
The superpotential $W+w$, given by eqs.~(\ref{m11},\ref{m13}) has an isolated
stationary point determined by eqs.~(\ref{m14}-\ref{m16}).
Thanks to eq.~(\ref{m16})
the $\phi_i$ obey, up to a rescaling, the commutation relations of the
generators of $SU(2)$. Moreover, as discussed in the text, they act on the
fundamental of $SU(N)$ as the irreducible representation of dimension $N$.
This means that one can define
\beq
\phi_i=-imL_i.
\eeq{c2}
$L_i$ acts on the adjoint of $SU(N)$, that is on the traceless product of two
fundamentals. In terms of representations of $SU(2)$, $L_i$ acts on the
reducible representation
\beq
j\times j - 0= 2j + (2j-1) +... + 1, \;\;\; j=(N-1)/2.
\eeq{c3}
Noticing that $i\epsilon_{ilj}$ acts on the index $i$ as the $j=1$
representation of $SU(2)$, we can re-write eq.~(\ref{c1}) in a remarkably
simple way:
\beq
(W+w)''= -m (S_iL_i +1).
\eeq{c4}
In this equation, $S_i$ are the generators of $SU(2)$ in the $j=1$
representation.

Eq.~(\ref{c4}) is easily diagonalized in terms of the eigenstates of the
``total angular momentum'' $\vec{J}^2=(\vec{S}+ \vec{L})^2$. The eigenvalues
and their multiplicity are given in Table 1.

\begin{table}[h]\centering
\caption{Eigenvalues of $(W+w)''$}
\begin{tabular}{|c|c|c|c|}
\hline $\vec{L}$ irrep & $\vec{J}$ irrep & Eigenvalue & Multiplicity \\
\hline\hline
 $1\leq l \leq 2j$ & $l+1 $ & $-m(l+1)$ & $2l + 3$ \\ \hline $l$ & $l$ & $0$
& $2l +1$ \\
\hline $l$ & $l-1$ & $ml$ & $2l -1$ \\ \hline
\end{tabular}
\end{table}

The zero eigenvalues of $(W+w)''$ are due to gauge invariance, while the
corresponding eigenvectors mix with the gauginos to give states with ``fermion
mass'' $\pm m$.

For our purpose, it is useful to choose $m$ to be real and
positive, and use gauge invariance to put $L_3$ in eq.~(\ref{c3}) in the form
\beq
L_3={\rm diag}\,(j,j-1,...,-j).
\eeq{c4a}
With this choice, the fields in the Cartan subalgebra of $SU(N)$ are diagonal,
up to an $SU(N)$ transformation. Also, at a generic point in moduli space, all
and only the fields in the Cartan subalgebra are light. This suggests to
introduce another basis for the chiral multiplets, labeled by the eigenvalues
of $L_3$ and $S_3$, $l_3,s_3$. In this basis the ``light'' fermions (i.e. the
slow modes in the adiabatic expansion) read
$U^\dagger \psi_{l_3=0,s_3,\alpha}U$, $U\in SU(N)$,
while the heavy fermions (fast
modes) read $U^\dagger \psi_{l_3\neq 0,s_3,\alpha}U$.
Now, we have two basis with which to label the fermions of the chiral
multiplet. One, labeled by the eigenvalues of $\vec{J}^2, J_3$ diagonalizes
the fermion mass matrix around the stationary point given by
eqs.~(\ref{m14}-\ref{m16}).
The other, labeled by the eigenvalues of $L_3,S_3$, diagonalizes the fermion
mass matrix along the moduli space. As shown in Appendix B,
the fermionic wave function, $F(\phi)$, is proportional to the Fock
vacuum determined by $N(0)$. In our case, this means, among other things, that
$F(\phi)$ obeys the following equation:
\beq
\alpha^{(l-1)}_{j_3}F(\phi)= 0 \Rightarrow
U^\dagger\alpha^{(l-1)}_{j_3}UF(\phi)= 0, \;\;\; -1\leq j_3 \leq 1.
\eeq{c5}
Here, we have used the gauge invariance of $F(\phi)$, and denoted by
$\alpha^{(l-1)}_{j_3}$ the fermion annihilation operator in the basis
$\vec{J}^2,J_3$, belonging to the $\vec{J}$ representation $(l-1)$.
When expressed in terms of the creators and annihilators
$\beta^\dagger_{l_3s_3}$, $\beta_{l_3,s_3}$, in the basis $L_3,S_3$,
this operator obeys:
\beq
U^\dagger\alpha^{(l-1)}_{j_3}U=\langle l-1,j_3|l,1,0,j_3\rangle
\beta^\dagger_{0,j_3} +
\sum_{m\neq 0}\langle l-1,j_3|l,1,m,j_3-m\rangle\beta^\#_{m,j_3-m}.
\eeq{c6}
Here $\beta^\#$ stands for either $\beta$ or $\beta^\dagger$, while
$\langle l-1,j_3|l,1,l_3,s_3\rangle$ are
Clebsch-Gordan coefficients.

Notice the presence of the creation operator
$\beta^\dagger_{0,j_3}$ on the r.h.s. of eq.~(\ref{c6}). This is a crucial
point that needs some explaining.
The matrix $(W+w)_{ij}$ in eq.~(\ref{c1}), computed at the stationary point
is real, since it is given by eq.~(\ref{c3}). This implies that if
$\psi_1$ and $\psi_2$ are two eigenvectors of $(W+w)''$, with eigenvalue
$\omega$, then the eigenstates of the ``fermion mass'' are
$\psi_\pm=\psi_1\pm \psi^\dagger_2$,
with eigenvalues $\pm \omega$. Along the valley given by
eq.~(\ref{m21}), $(W+w)''$ can also be diagonalized. Let us label by
$\tilde{\psi}^a_1$, $\tilde{\psi}^a_2$ a basis of eigenstates with
eigenvalues $\omega^a$. Expanding $\psi_\alpha$ in this new basis we find
$U^\dagger\psi_\alpha U=\sum_a c_a\tilde{\psi}^a_\alpha $, $U\in SU(N)$.
We also have:
$U^\dagger\psi_\pm U=\sum_a (c_a\tilde{\psi}^a_1 \pm c^*_a
\tilde{\psi}^{a\,\dagger}_2)$. When, say, $\omega^1$ and $c_1$ are real,
one may rewrite this equation in the form
$U^\dagger\psi_\pm U=c_1\tilde{\psi}^1_\pm + ... $.
In this last equation, $\psi_+$ and $\tilde{\psi}^1_+$ are both creators or
both destructors if  $\omega$ and $\omega^1$ have the same sign. When
$\omega$ and $\omega^1$ have opposite sign, instead, $\psi_+$ is a creator
when $\tilde{\psi}^1_+$ is a destructor, and vice versa.
In our case, $c_1$ is a Clebsch-Gordan
coefficient, which is indeed real; the eigenvalue $\omega$ is
$ml$, while $\omega^1$ is $-m$, which explains eq.~(\ref{c6}).

By writing eq.~(\ref{c5}) in terms of the oscillators $\beta$, we get an
equation that can be simplified by noticing that on the moduli space, away
from the origin $\phi_i=0$, the charged fermions created by
$\beta^\dagger_{m,s_3}$, $m\neq 0$ have a very large frequency
$\propto |\phi_i|$.
This implies that all components of $F(\phi)$, except the one proportional to
the ground state of these
oscillators, decay as fast as $\exp(-{\rm const}\,|\phi_i|^4$).
Eq.~(\ref{c5}) can thus be rewritten as
\beq
(\langle l-1,j_3|l,1,0,j_3\rangle\beta^\dagger_{0,j_3} +
\sum_{m\neq 0} c^m_{j_3}\beta^\dagger_{m,j_3-m})F(\phi)=0.
\eeq{c7}
Here $c^m_{j_3}=\langle l-1,j_3|l,1,m,j_3-m\rangle$, if
$\beta^\#_{m,j_3-m}= \beta^\dagger_{m,j_3-m}$,
and $c^m_{j_3}=0$, otherwise.
Notice that the sum in eq.~(\ref{c7}) contains at most 2 non-vanishing terms.
Since $F(\phi)$ has zero charged-fermion number, in the large-$\phi_i$
region of moduli space all terms in the sum have to vanish separately. Since
$\langle l-1,j_3|l,1,0,j_3\rangle$ vanishes only for $j_3=\pm l$,
this may only happen when all $c^m_{j_3}$ vanish, and
$\beta^\dagger_{0,j_3}F(\phi)=0$ for all $|j_3|\leq \min[1,l-1]$.
This equation means that $F(\phi)$, is
the completely filled state in the Fock space of the oscillators
$\beta^\#_{0,j_3}$, $|j_3|\leq \min[1,l-1]$.

\end{document}